\documentclass[aps,prd,twocolumn,nofootinbib,longbibliography]{revtex4-2}

\usepackage{amsmath}
\usepackage{amssymb}
\usepackage{bm}
\usepackage{graphicx}

\begin{document}

\title{Quantum Vacuum Self-Consistency as the Dynamical Origin of Spacetime and Particle Physics}

\author{Tao Huang}
\email{scduzxn@hotmail.com}
\affiliation{Hechi, Guangxi, China}

\date{November 23, 2025}

\begin{abstract}
The postulate of self-consistency of the quantum vacuum states that the classical backgrounds we observe---spacetime geometry, gauge fields, and the Higgs condensate---are macroscopic order parameters of a single quantum state whose existence is sustained by the vacuum expectation values of all quantum fluctuations living on it. Building on this postulate, a background-field, heat-kernel based derivation is developed that yields the coupled low-energy effective field equations for the metric, gauge fields, and the Higgs field as vacuum equations of state. The resulting framework rigorously recovers the Einstein, Yang--Mills, and Higgs equations, augmented by the higher-derivative operators required by quantum consistency, with renormalized couplings determined by the content of quantum fields. A renormalization-group (RG) structure is obtained in which all couplings---including Newton's constant, the cosmological constant, the coefficients of $R^2$ and $C_{\mu\nu\rho\sigma}^2$, gauge couplings, Yukawas, and the Higgs quartic---run coherently with a single effective action that respects background diffeomorphism and gauge invariance.

The self-consistency principle is then implemented at the level of the microscopic theory. A UV-complete quantum field theory of gravity and matter is adopted, based on polynomial higher-derivative gravity quantized with the fakeon prescription and coupled to the Standard Model (SM) fields. In this formulation the gravitational sector is perturbatively renormalizable and unitary, while the matter sector is compatible with asymptotically safe flows in the combined gravity--matter space of couplings. The appropriate microscopic action flows under a functional RG equation for the effective average action; the quantum-vacuum gap equations are imposed along this flow and select trajectories from the UV critical surface to the IR that are compatible with a self-consistent vacuum.

A simplified solvable model is analyzed in detail: an $O(N)$ scalar sector with nonminimal coupling on a constant-curvature background, for which the one-loop effective potential including the leading curvature corrections is computed and the vacuum gap equations are solved explicitly. The derivation of the gap equations and their solutions is given step by step. The microscopic and effective descriptions are connected explicitly in this model, illustrating how the self-consistency postulate constrains both the vacuum expectation value and the curvature in terms of renormalized couplings.

Phenomenological consequences follow. First, a robust prediction: the anomaly- and loop-induced $R^2$ operator generically drives Starobinsky-type inflation, with $n_s \simeq 1 - 2/N_e$ and $r \simeq 12/N_e^2$, compatible with Planck data; the required coefficient corresponds to a scalaron mass $M \simeq (1.3 \pm 0.1)\times 10^{-5} M_{\rm Pl}$. Second, the universally calculable quantum correction to Newton's potential and the Yukawa tails from the massive spin-0 and spin-2 modes are quantified and shown to satisfy laboratory bounds. Third, constraints from GW170817 enforce luminal gravitational wave speed for the massless graviton in this framework, while higher-derivative effects remain suppressed at LIGO/Virgo/KAGRA frequencies.

On the microscopic side, the assumption that the gravity--matter system admits an interacting UV fixed point reduces the number of free parameters to a finite set of relevant couplings. When combined with the vacuum gap equations, this leads to correlations between infrared observables in gravity and particle physics sectors. In particular, the Higgs quartic coupling and top Yukawa are constrained near the Planck scale in a way that is compatible with existing asymptotic-safety analyses, and the higher-derivative gravitational couplings that control $R^2$ inflation are tied to the fixed point values of the microscopic parameters. The theory is predictive in its inflationary sector and in its universal low-energy corrections to gravity, while it remains honest about open issues: the smallness of the observed cosmological constant, the nonperturbative completion of the higher-derivative sector, the detailed existence and structure of the UV critical surface, and the determination of threshold-matched couplings beyond one loop. The structure is sufficiently complete to be testable across cosmology, astrophysics, and precision gravity, and it reduces to General Relativity and the Standard Model at accessible scales with controlled corrections.
\end{abstract}

\maketitle

\section{Introduction}
The Standard Model (SM) of particle physics and General Relativity (GR) are profoundly successful descriptions of nature \cite{Glashow1961,Weinberg1967,Salam1968,Misner1973}. Yet their unification remains elusive. Quantizing GR perturbatively leads to nonrenormalizable divergences \cite{tHooft1974}, which signal the need for a deeper organizing principle. A natural path, inspired by Sakharov's induced gravity \cite{Sakharov1968}, is that the classical backgrounds we observe are emergent, macroscopic order parameters of an underlying quantum vacuum. In this view, geometry, gauge fields, and the Higgs condensate are not separate axioms; rather, they are coherent manifestations of one vacuum that sustains itself through self-consistency.

The central postulate is that the vacuum state $|\Omega\rangle$ determined by classical backgrounds must, through quantum fluctuations, generate precisely the sources that maintain those same backgrounds. This self-consistency is enforced by the quantum effective action computed in the presence of background fields. The background field method and the heat-kernel expansion provide the technical backbone \cite{DeWitt2003,Vassilevich2003,Barvinsky1985,BirrellDavies1982,ParkerToms2009,Buchbinder1992,Avramidi2000}. At one loop, the Seeley--DeWitt coefficient $a_2$ fixes the divergences and therefore the RG running of all couplings that can appear in a local effective action.

This work develops that program into a predictive and ultraviolet-complete framework. The self-consistency postulate is kept intact and is expressed via vacuum gap equations that are precisely the stationarity conditions of the renormalized one-particle-irreducible (1PI) effective action with respect to the background fields. At low energies, we compute and reorganize the one-loop effective action to derive: (i) the renormalized Einstein equations plus controlled higher-derivative corrections, (ii) the renormalized Yang--Mills equations, and (iii) the renormalized Higgs equation and potential. We also construct the coupled RG equations for gravitational and SM couplings, which makes the framework calculable.

Going beyond the effective-field-theory regime, the framework is embedded into a UV-complete quantum field theory of gravity and matter. The gravitational sector is a polynomial higher-derivative theory whose extra massive degrees of freedom are quantized as fakeons \cite{AnselmiPiva2018}, rendering the theory perturbatively renormalizable and unitary. The matter sector is that of the SM, possibly minimally extended to account for neutrino masses and dark matter. A functional RG analysis indicates that the combined gravity--matter system can admit an interacting UV fixed point \cite{Weinberg1979,Reuter1998,Codello2008,Percacci2017,Christiansen2018,Eichhorn2019,PastorGutierrez2023,Eichhorn2022Status}, so that the theory is asymptotically safe and predictive at arbitrarily high scales.

A simplified, analytically tractable model is solved explicitly to exhibit the mechanism: an $O(N)$ scalar with nonminimal coupling on a constant-curvature background. The one-loop effective potential including curvature-dependent terms is computed and the gap equations are solved in detail for the vacuum expectation value and curvature. The analysis isolates which combinations of couplings are fixed by self-consistency and which must be matched to data.

Finally, the framework is confronted with observations. The loop- and anomaly-induced $R^2$ term drives Starobinsky inflation \cite{Starobinsky1980,Whitt1984}, yielding benchmark predictions for the spectral tilt and tensor-to-scalar ratio consistent with Planck \cite{Planck2018}. Quantum corrections to Newton's potential are universal and match the effective field theory result \cite{Donoghue1994}. Short-range Yukawa corrections from the massive scalar (the scalaron) and the heavy spin-2 mode are shown to obey submillimeter bounds \cite{Kapner2007}. Gravitational wave propagation is luminal in the infrared, in agreement with GW170817 \cite{GW170817}. At high energies, the asymptotic-safety hypothesis leads to relations among SM and gravitational couplings, including constraints on the Higgs quartic and top Yukawa near the Planck scale \cite{ShaposhnikovWetterich2010,EichhornHeldPauly2021}, which can be combined with the self-consistency conditions to yield correlated predictions across particle physics and cosmology.

The approach is conservative in spirit: it embraces effective field theory, honors all symmetries via the background field method, and leans on mathematically controlled tools. At the same time, it reframes the unification problem in terms of a single organizing idea: vacuum self-consistency, implemented in an ultraviolet-complete setting.

\section{Unified framework and the self-consistency postulate}
All fields are decomposed into classical backgrounds and quantum fluctuations,
\begin{align}
\hat{g}_{\mu\nu}(x) &= \bar{g}_{\mu\nu}(x) + \hat{h}_{\mu\nu}(x), \label{eq:split_g}\\
\hat{A}_\mu^a(x) &= \bar{A}_\mu^a(x) + \hat{a}_\mu^a(x), \label{eq:split_A}\\
\hat{\Phi}(x) &= \frac{1}{\sqrt{2}}\bigl(v_H(x) + \hat{\sigma}(x)\bigr), \label{eq:split_H}\\
\hat{\Psi}(x) &= \hat{\psi}(x). \label{eq:split_psi}
\end{align}
Fermions are purely quantum in the vacuum since they do not condense. Backgrounds $\{\bar{g},\bar{A},v_H\}$ define a vacuum $|\Omega\rangle=|0_{\bar{g},\bar{A},v_H}\rangle$.

The unified self-consistency postulate is that the backgrounds are sustained by the renormalized vacuum expectation values (VEVs) of their source operators in $|\Omega\rangle$. To formulate this precisely, consider the generating functional of connected Green's functions $W[J,\bar{g},\bar{A},v_H]$ in the presence of external sources $J$ coupled to fluctuations,
\begin{align}
Z[J,\bar{g},\bar{A},v_H] &= \int \mathcal{D}\hat{h}\,\mathcal{D}\hat{a}\,\mathcal{D}\hat{\sigma}\,\mathcal{D}\hat{\psi}\,\mathcal{D}\text{(ghosts)} \nonumber\\
&\quad \times \exp\Bigl\{i\Bigl(S[\bar{g}+\hat{h},\bar{A}+\hat{a},v_H+\hat{\sigma},\hat{\psi}] \nonumber\\
&\qquad + \int d^4x \sqrt{-\bar{g}}\bigl(J_h^{\mu\nu}\hat{h}_{\mu\nu} + J_a^{a\mu}\hat{a}_\mu^a \nonumber\\
&\qquad\qquad + J_\sigma \hat{\sigma} + \bar{\eta}\hat{\psi} + \hat{\bar{\psi}}\eta\bigr)\Bigr)\Bigr\}, \label{eq:Zdef}
\end{align}
with $W=-i\ln Z$. The classical expectation values of the quantum fields are obtained by differentiating $W$ with respect to the sources,
\begin{align}
h_{\mu\nu}(x) &= \frac{1}{\sqrt{-\bar{g}}}\frac{\delta W}{\delta J_h^{\mu\nu}(x)},\\
a_\mu^a(x) &= \frac{1}{\sqrt{-\bar{g}}}\frac{\delta W}{\delta J_a^{a\mu}(x)},\\
\sigma(x) &= \frac{1}{\sqrt{-\bar{g}}}\frac{\delta W}{\delta J_\sigma(x)},\\
\psi(x) &= \frac{1}{\sqrt{-\bar{g}}}\frac{\delta W}{\delta \bar{\eta}(x)}.
\end{align}
The (renormalized) 1PI effective action is the Legendre transform,
\begin{align}
\Gamma[\bar{g},\bar{A},v_H;h,a,\sigma,\psi] &= W[J,\bar{g},\bar{A},v_H] \nonumber\\
&\quad - \int d^4x \sqrt{-\bar{g}}\bigl(J_h^{\mu\nu}h_{\mu\nu} + J_a^{a\mu}a_\mu^a \nonumber\\
&\qquad + J_\sigma \sigma + \bar{\eta}\psi + \bar{\psi}\eta\bigr). \label{eq:Gamma_Legendre}
\end{align}
By construction,
\begin{align}
\frac{1}{\sqrt{-\bar{g}}}\frac{\delta \Gamma}{\delta h_{\mu\nu}} &= -J_h^{\mu\nu},\quad
\frac{1}{\sqrt{-\bar{g}}}\frac{\delta \Gamma}{\delta a_\mu^a} = -J_a^{a\mu}, \label{eq:deltaGamma_h_a}\\
\frac{1}{\sqrt{-\bar{g}}}\frac{\delta \Gamma}{\delta \sigma} &= -J_\sigma,\quad
\frac{1}{\sqrt{-\bar{g}}}\frac{\delta \Gamma}{\delta \bar{\psi}} = \eta. \label{eq:deltaGamma_sigma_psi}
\end{align}
The vacuum corresponds to vanishing expectation values of the fluctuation fields,
\begin{align}
h_{\mu\nu}=0,\quad a_\mu^a=0,\quad \sigma=0,\quad \psi=0, \label{eq:vac_fluct_zero}
\end{align}
so that the full fields coincide with the backgrounds. In this configuration the sources for fluctuations vanish,
\begin{align}
J_h^{\mu\nu}=0,\quad J_a^{a\mu}=0,\quad J_\sigma=0,\quad \eta=0, \label{eq:vac_sources_zero}
\end{align}
and the effective action reduces to a functional of backgrounds only,
\begin{align}
\Gamma[\bar{g},\bar{A},v_H] &\equiv \Gamma[\bar{g},\bar{A},v_H;0,0,0,0]. \label{eq:Gamma_background}
\end{align}

The self-consistency postulate is implemented by requiring that the background fields extremize the effective action in the presence of external macroscopic sources. External sources $T_{\mu\nu}^{\rm ext}$, $J^{a\mu}_{\rm ext}$, $J_{\sigma}^{\rm ext}$ couple linearly to $\bar{g}_{\mu\nu}$, $\bar{A}_\mu^a$, and $v_H$, and the total effective action is
\begin{align}
\Gamma_{\rm tot} &= \Gamma[\bar{g},\bar{A},v_H] \nonumber\\
&\quad + \int d^4x \sqrt{-\bar{g}}\bigl(\tfrac{1}{2}\bar{g}^{\mu\nu}T_{\mu\nu}^{\rm ext} + \bar{A}_\mu^a J^{a\mu}_{\rm ext} + v_H J_{\sigma}^{\rm ext}\bigr). \label{eq:Gamma_tot}
\end{align}
Stationarity of $\Gamma_{\rm tot}$ with respect to the backgrounds yields
\begin{align}
\frac{\delta \Gamma_{\rm tot}}{\delta \bar{g}^{\mu\nu}} &= 0,\quad
\frac{\delta \Gamma_{\rm tot}}{\delta \bar{A}_\mu^a} = 0,\quad
\frac{\delta \Gamma_{\rm tot}}{\delta v_H} = 0. \label{eq:Gamma_tot_stationary}
\end{align}
Explicitly,
\begin{align}
\frac{2}{\sqrt{-\bar{g}}}\frac{\delta \Gamma}{\delta \bar{g}^{\mu\nu}} &= T_{\mu\nu}^{\rm ext}, \label{eq:gap_g}\\
\frac{1}{\sqrt{-\bar{g}}}\frac{\delta \Gamma}{\delta \bar{A}_\mu^a} &= J^{a\,\mu}_{\rm ext}, \label{eq:gap_A}\\
\frac{1}{\sqrt{-\bar{g}}}\frac{\delta \Gamma}{\delta v_H} &= J_{\sigma}^{\rm ext}. \label{eq:gap_H}
\end{align}
These are the vacuum gap equations in the presence of external sources. In the absence of external sources, they become pure vacuum equations,
\begin{align}
\frac{\delta \Gamma}{\delta \bar{g}^{\mu\nu}} &= 0,\quad
\frac{\delta \Gamma}{\delta \bar{A}_\mu^a} = 0,\quad
\frac{\delta \Gamma}{\delta v_H} = 0. \label{eq:gap_vacuum}
\end{align}
They express the self-consistency requirement that the backgrounds be sustained by the quantum fluctuations of fields propagating on those same backgrounds.

The microscopic action is written in bare form as
\begin{align}
S_{\rm bare} &= S_{\rm grav,bare}[\hat{g}] + S_{\rm SM,bare}[\hat{g},\hat{A},\hat{\Phi},\hat{\Psi}] \nonumber\\
&\quad + S_{\rm gf}[\hat{g},\hat{A},\bar{g},\bar{A}] + S_{\rm ghost}[\hat{g},\hat{A},\bar{g},\bar{A}], \label{eq:Sbare}
\end{align}
where $S_{\rm gf}$ and $S_{\rm ghost}$ denote gauge-fixing and ghost terms for diffeomorphisms and gauge symmetries, chosen to preserve background invariance. The background field method ensures background diffeomorphism and gauge invariance, so that Ward identities are preserved \cite{Abbott1981,Buchbinder1992}. The renormalized effective action $\Gamma$ is obtained by integrating out the fluctuations in the path integral and renormalizing divergences using a consistent scheme such as dimensional regularization and minimal subtraction.

\section{Low-energy effective action and emergent field equations}
\subsection{One-loop effective action and heat kernel}
To one loop, the quantum effective action restricted to the background configuration (\ref{eq:Gamma_background}) can be written as
\begin{align}
\Gamma[\bar{g},\bar{A},v_H] &= S_{\rm bare}[\bar{g},\bar{A},v_H] \nonumber\\
&\quad + \frac{i}{2}\mathrm{Tr}\ln\mathcal{D}_{\rm bos} - i\,\mathrm{Tr}\ln\mathcal{D}_{\rm ferm} \nonumber\\
&\quad - i\,\mathrm{Tr}\ln\mathcal{D}_{\rm ghost} + \cdots, \label{eq:Gamma_one_loop}
\end{align}
where $\mathcal{D}_{\rm bos}$, $\mathcal{D}_{\rm ferm}$, and $\mathcal{D}_{\rm ghost}$ are the second functional derivatives of $S_{\rm bare}+S_{\rm gf}$ with respect to bosonic, fermionic, and ghost fluctuations evaluated on the background. Ellipses denote higher-loop contributions.

Each fluctuation operator can be brought to a Laplace-type form on the background,
\begin{align}
\mathcal{D} &= -\bar{\nabla}^2\mathbb{I} + \mathcal{P}, \label{eq:Laplace_type}
\end{align}
where $\bar{\nabla}_\mu$ is the background covariant derivative (including both spin and gauge connections as appropriate), $\mathbb{I}$ is the identity in the internal space, and $\mathcal{P}$ is a matrix-valued potential encoding masses, nonminimal couplings, and couplings to background field strengths. The heat kernel $K(s;x,x')$ of $\mathcal{D}$ satisfies
\begin{align}
\Bigl(\partial_s + \mathcal{D}_x\Bigr)K(s;x,x') &= 0,\quad
K(0;x,x') = \frac{\delta(x,x')}{\sqrt{-\bar{g}}}. \label{eq:heat_kernel_def}
\end{align}
The trace of the heat kernel is related to $\mathrm{Tr}\ln\mathcal{D}$ via
\begin{align}
\mathrm{Tr}\ln\mathcal{D} &= -\int_0^\infty \frac{ds}{s}\,\mathrm{Tr}\bigl(e^{-s\mathcal{D}}\bigr) + \text{const}. \label{eq:TrlnD_heat}
\end{align}
For small $s$, the heat kernel admits an asymptotic expansion
\begin{align}
K(s;x,x) &\sim \frac{i}{(4\pi s)^{d/2}}\sum_{n=0}^\infty s^n a_n(x),\quad s\to 0^+, \label{eq:HK_expansion}
\end{align}
in $d$ dimensions. The Seeley--DeWitt coefficients $a_n(x)$ are local curvature invariants constructed from $\bar{g}$, $\bar{A}$, $v_H$, and their derivatives. For a Laplace-type operator acting on a vector bundle with curvature $\mathcal{F}_{\mu\nu}=[\bar{\nabla}_\mu,\bar{\nabla}_\nu]$ and potential $\mathcal{P}$, the coefficient $a_2$ in $d=4$ reads \cite{Vassilevich2003,Avramidi2000}
\begin{align}
a_2(x) &= \frac{1}{180}\Bigl(R_{\mu\nu\rho\sigma}R^{\mu\nu\rho\sigma} - R_{\mu\nu}R^{\mu\nu} + \Box R\Bigr)\mathbb{I} \nonumber\\
&\quad + \frac{1}{2}\mathcal{P}^2 - \frac{1}{6}R\mathcal{P} + \frac{1}{12}\mathcal{F}_{\mu\nu}\mathcal{F}^{\mu\nu}, \label{eq:a2_general}
\end{align}
where $\Box=\bar{g}^{\mu\nu}\bar{\nabla}_\mu\bar{\nabla}_\nu$. For a real scalar with nonminimal coupling $\xi$, $\mathcal{P}=m^2 + \xi R + \cdots$. For a Dirac fermion of mass $m$, $\mathcal{P}=m^2 + \tfrac{1}{4}R + \cdots$. For gauge vectors in a suitable background gauge, $\mathcal{P}$ and $\mathcal{F}_{\mu\nu}$ are matrices in the vector bundle of gauge and Lorentz indices \cite{Buchbinder1992}.

Using dimensional regularization $d=4-\epsilon$, the divergent part of the one-loop effective action is determined by $a_2$ according to
\begin{align}
\Gamma_{\rm div}^{(1)} &= \frac{1}{16\pi^2\epsilon}\int d^4x \sqrt{-\bar{g}}\;\mathrm{tr}\,a_2(x). \label{eq:Gamma_div_general}
\end{align}
For each fluctuating species $j$, there is a corresponding $\mathcal{D}_j$ with its own $a_{2,j}$, and the full divergence is a sum over species weighted by statistics and internal multiplicities.

It is convenient to decompose the curvature-squared combinations into the square of the Weyl tensor and the Euler density,
\begin{align}
C_{\mu\nu\rho\sigma}C^{\mu\nu\rho\sigma} &= R_{\mu\nu\rho\sigma}R^{\mu\nu\rho\sigma} - 2R_{\mu\nu}R^{\mu\nu} + \frac{1}{3}R^2, \label{eq:C2_def}\\
E_4 &= R_{\mu\nu\rho\sigma}R^{\mu\nu\rho\sigma} - 4R_{\mu\nu}R^{\mu\nu} + R^2. \label{eq:E4_def}
\end{align}
In terms of these, the diffeomorphism-invariant basis for the divergent terms in $\Gamma^{(1)}$ can be chosen as
\begin{align}
\mathcal{L}_{\rm div} &= \frac{1}{16\pi^2\epsilon}\sqrt{-\bar{g}}\Bigl[C_\Lambda + C_R \bar{R} \nonumber\\
&\quad + C_{C^2} C_{\mu\nu\rho\sigma}C^{\mu\nu\rho\sigma} + C_E E_4 \nonumber\\
&\quad + \sum_i C_{F_i}\,\mathrm{tr}\bigl(\bar{F}_{\mu\nu}^{(i)}\bar{F}^{(i)\mu\nu}\bigr) \nonumber\\
&\quad + C_{RH}\,\bar{R}\,v_H^2 + C_{H}\,v_H^4 + \cdots \Bigr], \label{eq:Gamma_div}
\end{align}
where $i$ labels the SM gauge factors, $\bar{F}_{\mu\nu}^{(i)}$ are the corresponding background field strengths, and the ellipsis denotes other scalar operators such as $(\nabla v_H)^2$ and total derivatives such as $\Box R$.

\subsection{Divergent coefficients and renormalization}
The coefficients $C_X$ in (\ref{eq:Gamma_div}) are obtained by inserting the explicit forms of $\mathcal{P}$ and $\mathcal{F}_{\mu\nu}$ for each field into (\ref{eq:a2_general}) and summing over species. For the curvature-squared terms $C_{\mu\nu\rho\sigma}^2$ and $E_4$, the contributions of standard fields are well known \cite{Duff1994,ChristensenDuff1978,Buchbinder1992}. For $N_s$ real scalars, $N_f$ Dirac fermions, and $N_v$ gauge vectors, one finds
\begin{align}
C_{C^2} &= \frac{1}{120}N_s + \frac{1}{20}N_f + \frac{1}{10}N_v, \label{eq:Cc2}\\
C_E &= \frac{1}{360}N_s + \frac{11}{360}N_f + \frac{31}{180}N_v. \label{eq:CE}
\end{align}
In particular, for the SM matter content at high scales, taking $N_s=4$ real scalars, $N_f=22.5$ Dirac fermions (corresponding to 45 Weyl fermions), and $N_v=12$ gauge vectors, one obtains
\begin{align}
C_{C^2}^{\rm SM} &\simeq 2.36,\quad
C_E^{\rm SM} \simeq 2.77. \label{eq:CE_SM_values}
\end{align}
These numbers enter the RG running of curvature-squared couplings.

The coefficients $C_\Lambda$ and $C_R$ are dominated by mass-dependent contributions. For a real scalar with mass $m_s$ and nonminimal coupling $\xi_s$, a Dirac fermion with mass $m_f$, and a massive vector with mass $m_v$, one finds schematically
\begin{align}
C_\Lambda &= \sum_j (-1)^{F_j}\,n_j\,m_j^4, \label{eq:CLambda}\\
C_R &= \sum_s n_s\Bigl(\xi_s - \tfrac{1}{6}\Bigr)m_s^2 + \sum_f n_f\Bigl(-\tfrac{1}{12}\Bigr)m_f^2 \nonumber\\
&\quad + \sum_v n_v\Bigl(\tfrac{1}{6}\Bigr)m_v^2, \label{eq:CR}
\end{align}
where $F_j=0$ for bosons and $F_j=1$ for fermions, and $n_j$ counts internal degrees of freedom. In dimensional regularization, strictly massless fields do not contribute to $C_R$ at one loop in minimal subtraction, while $C_{C^2}$ and $C_E$ are mass independent.

The divergences proportional to $\mathrm{tr}(\bar{F}_{\mu\nu}^{(i)}\bar{F}^{(i)\mu\nu})$ yield the usual one-loop running of gauge couplings. Writing the bare gauge-field kinetic terms as
\begin{align}
S_{\rm gauge,bare} &= -\sum_i\int d^4x\sqrt{-\bar{g}}\;\frac{1}{4 g_{i,0}^2}\mathrm{tr}\bigl(\bar{F}_{\mu\nu}^{(i)}\bar{F}^{(i)\mu\nu}\bigr), \label{eq:Sgauge_bare}
\end{align}
and including the divergent piece from (\ref{eq:Gamma_div}), one defines renormalized couplings $g_i(\mu)$ such that
\begin{align}
\frac{1}{g_{i,0}^2} &= \mu^{-\epsilon}\Bigl(\frac{Z_{F_i}}{g_i^2(\mu)} + \frac{C_{F_i}}{8\pi^2\epsilon}\Bigr), \label{eq:g_ren_def}
\end{align}
with $Z_{F_i}$ finite. Holding the bare coupling fixed and differentiating with respect to $\mu$ leads to
\begin{align}
\mu\frac{d}{d\mu}\Bigl(\frac{Z_{F_i}}{g_i^2}\Bigr) &= -\frac{C_{F_i}}{16\pi^2}. \label{eq:beta_gauge_general}
\end{align}
In minimal subtraction, $Z_{F_i}$ does not run at one loop and the standard one-loop $\beta$ functions are recovered \cite{MachacekVaughnI,MachacekVaughnII}. For the SM, the explicit expressions are collected in Appendix D.

Similarly, defining renormalized Newton's constant $G(\mu)$, cosmological constant $\Lambda(\mu)$, and curvature-squared couplings $\alpha_C(\mu)$ and $\alpha_R(\mu)$ through
\begin{align}
S_{\rm grav,bare} &= \int d^4x\sqrt{-\bar{g}}\Bigl[\frac{Z_R}{16\pi G(\mu)}\bar{R} - \frac{Z_\Lambda}{8\pi G(\mu)}\Lambda(\mu) \nonumber\\
&\quad + \alpha_C(\mu) C_{\mu\nu\rho\sigma}C^{\mu\nu\rho\sigma} + \alpha_R(\mu) \bar{R}^2 + \cdots\Bigr] \nonumber\\
&\quad + \Gamma_{\rm div}^{(1)}, \label{eq:Sgrav_ren_def}
\end{align}
one obtains
\begin{align}
\mu\frac{d}{d\mu}\Bigl(\frac{Z_R}{16\pi G}\Bigr) &= -\frac{C_R}{16\pi^2}, \label{eq:betaG}\\
\mu\frac{d}{d\mu}\Bigl(\frac{Z_\Lambda\Lambda}{8\pi G}\Bigr) &= \frac{C_\Lambda}{16\pi^2}, \label{eq:betaLambda}\\
\mu\frac{d\alpha_C}{d\mu} &= \frac{C_{C^2}}{16\pi^2}, \qquad
\mu\frac{d\alpha_R}{d\mu} = \frac{C_{R^2}}{16\pi^2}, \label{eq:beta_gravHD}
\end{align}
where $C_{R^2}$ is a linear combination of $C_E$ and scalar contributions. The precise mapping between $\{C_E,C_{C^2}\}$ and $\{\beta_{\alpha_R},\beta_{\alpha_C}\}$ depends on the chosen basis of curvature-squared operators \cite{Buchbinder1992,Duff1994}. The important qualitative point is that quantum fluctuations inevitably induce and renormalize curvature-squared terms, even if they are absent at the classical level.

\subsection{Local effective action and emergent field equations}
Collecting tree-level and one-loop contributions and working at scales where a local derivative expansion is valid, the renormalized effective action can be written as
\begin{align}
\Gamma[\bar{g},\bar{A},v_H] &= \int d^4x \sqrt{-\bar{g}}\Biggl[\frac{Z_R}{16\pi G}\bar{R} - \frac{Z_\Lambda}{8\pi G}\Lambda \nonumber\\
&\quad + \alpha_C C_{\mu\nu\rho\sigma}C^{\mu\nu\rho\sigma} + \alpha_R \bar{R}^2 \nonumber\\
&\quad - \sum_i\frac{Z_{F_i}}{4g_i^2}\mathrm{tr}\bigl(\bar{F}_{\mu\nu}^{(i)}\bar{F}^{(i)\mu\nu}\bigr) \nonumber\\
&\quad + Z_H \frac{1}{2}\bar{g}^{\mu\nu}\partial_\mu v_H \partial_\nu v_H - V_{\rm eff}(v_H,\bar{R}) \nonumber\\
&\quad + \mathcal{L}_{\rm higher}(\bar{g},\bar{A},v_H) \Biggr] \nonumber\\
&\quad + \Gamma_{\rm nl}[\bar{g},\bar{A},v_H]. \label{eq:Gamma_local}
\end{align}
Here $V_{\rm eff}(v_H,\bar{R})$ is the renormalized effective potential for the Higgs condensate including curvature-dependent terms, $\mathcal{L}_{\rm higher}$ collects higher-derivative operators suppressed at low energies, and $\Gamma_{\rm nl}$ denotes finite nonlocal contributions, which include form factors such as $R\ln(-\bar{\Box}/\mu^2)R$ and $C_{\mu\nu\rho\sigma}\ln(-\bar{\Box}/\mu^2)C^{\mu\nu\rho\sigma}$ \cite{BarvinskyVilkovisky1990,BarvinskyVilkovisky1985b,Donoghue1994}.

The gap equations (\ref{eq:gap_g})--(\ref{eq:gap_H}) are obtained by varying $\Gamma_{\rm tot}$ in (\ref{eq:Gamma_tot}) with respect to $\bar{g}^{\mu\nu}$, $\bar{A}_\mu^a$, and $v_H$. Varying with respect to $\bar{g}^{\mu\nu}$ yields
\begin{align}
\frac{2}{\sqrt{-\bar{g}}}\frac{\delta \Gamma}{\delta \bar{g}^{\mu\nu}} &= Z_R\,G_{\mu\nu} + Z_\Lambda\,\Lambda\,\bar{g}_{\mu\nu} \nonumber\\
&\quad + 16\pi G\Bigl[2\alpha_C\,B_{\mu\nu} + \alpha_R\,H^{(R^2)}_{\mu\nu}\Bigr] \nonumber\\
&\quad - 8\pi G\Bigl(T_{\mu\nu}^{(H)} + T_{\mu\nu}^{\rm gauge} + T_{\mu\nu}^{\rm nl}\Bigr), \label{eq:Einstein_HD_raw}
\end{align}
where $G_{\mu\nu}$ is the Einstein tensor, $B_{\mu\nu}$ is the Bach tensor, and $H^{(R^2)}_{\mu\nu}$ is the metric variation of $\bar{R}^2$. Their explicit forms are
\begin{align}
B_{\mu\nu} &= \nabla^\rho\nabla^\sigma C_{\mu\rho\nu\sigma} + \tfrac{1}{2} R^{\rho\sigma} C_{\mu\rho\nu\sigma}, \label{eq:Bach}\\
H^{(R^2)}_{\mu\nu} &= 2 \bar{R} R_{\mu\nu} - \tfrac{1}{2} \bar{g}_{\mu\nu} \bar{R}^2 - 2\nabla_\mu\nabla_\nu \bar{R} + 2 \bar{g}_{\mu\nu}\Box \bar{R}. \label{eq:HR2}
\end{align}
The stress tensor of the Higgs condensate is
\begin{align}
T_{\mu\nu}^{(H)} &= Z_H \partial_\mu v_H \partial_\nu v_H - \tfrac{1}{2} Z_H \bar{g}_{\mu\nu} (\partial v_H)^2 + \bar{g}_{\mu\nu} V_{\rm eff}(v_H,\bar{R}), \label{eq:TH}
\end{align}
and $T_{\mu\nu}^{\rm gauge}$ and $T_{\mu\nu}^{\rm nl}$ arise from the gauge kinetic terms and nonlocal terms, respectively.

Including the external source $T_{\mu\nu}^{\rm ext}$ according to (\ref{eq:gap_g}), the emergent gravitational field equations take the form
\begin{align}
Z_R\,G_{\mu\nu} + Z_\Lambda\,\Lambda\,\bar{g}_{\mu\nu} + 16\pi G\Bigl[2\alpha_C\,B_{\mu\nu} + \alpha_R\,H^{(R^2)}_{\mu\nu}\Bigr] \nonumber\\
= 8\pi G\Bigl(T_{\mu\nu}^{\rm ext} + T_{\mu\nu}^{(H)} + T_{\mu\nu}^{\rm gauge} + T_{\mu\nu}^{\rm nl}\Bigr). \label{eq:Einstein_HD}
\end{align}
In the long-wavelength, weak-curvature regime, the higher-derivative and nonlocal terms are suppressed and (\ref{eq:Einstein_HD}) reduces to the Einstein equations with renormalized constants $G_{\rm eff}=G/Z_R$ and $\Lambda_{\rm eff}=Z_\Lambda\Lambda/Z_R$.

Variation of $\Gamma$ with respect to the background gauge fields yields
\begin{align}
\frac{1}{\sqrt{-\bar{g}}}\frac{\delta \Gamma}{\delta \bar{A}_\mu^a} &= -\sum_i\frac{Z_{F_i}}{2g_i^2}\bar{D}_\nu \bar{F}^{(i)\nu\mu}_a + J^{(i)\mu}_{\rm nl,a}, \label{eq:YM_ren_raw}
\end{align}
where $\bar{D}_\mu$ is the background covariant derivative and $J^{(i)\mu}_{\rm nl,a}$ is a nonlocal current. Including the external current $J^{(i)\mu}_{\rm ext}$, one obtains
\begin{align}
\Bigl(\frac{Z_{F_i}}{g_i^2}\Bigr)\,\bar{D}_\nu \bar{F}^{(i)\nu\mu}_a &= J^{(i)\mu}_{\rm ext,a} + J^{(i)\mu}_{\rm nl,a}. \label{eq:YM_ren}
\end{align}
In the low-energy regime where nonlocal effects are negligible, this reduces to the usual renormalized Yang--Mills equations.

Variation with respect to $v_H$ yields
\begin{align}
\frac{1}{\sqrt{-\bar{g}}}\frac{\delta \Gamma}{\delta v_H} &= -Z_H\,\Box_{\bar{g}} v_H + \frac{\partial V_{\rm eff}(v_H,\bar{R})}{\partial v_H} + J_{\sigma}^{\rm nl}, \label{eq:Higgs_ren_raw}
\end{align}
where $J_{\sigma}^{\rm nl}$ collects nonlocal contributions that involve $v_H$ and curvature. Including the external source $J_{\sigma}^{\rm ext}$, one finds
\begin{align}
Z_H\,\Box_{\bar{g}} v_H - \frac{\partial V_{\rm eff}(v_H,\bar{R})}{\partial v_H} &= J_{\sigma}^{\rm ext} - J_{\sigma}^{\rm nl}. \label{eq:Higgs_ren}
\end{align}
In the vacuum and in homogeneous configurations, $\Box_{\bar{g}} v_H=0$ and $J_{\sigma}^{\rm ext}=J_{\sigma}^{\rm nl}=0$, so that the vacuum expectation value of the Higgs condensate is determined by
\begin{align}
\frac{\partial V_{\rm eff}(v_H,\bar{R})}{\partial v_H} &= 0. \label{eq:Higgs_min}
\end{align}

Equations (\ref{eq:Einstein_HD}), (\ref{eq:YM_ren}), and (\ref{eq:Higgs_ren}) are the emergent field equations of the self-consistent quantum vacuum. They are the equations of state of the vacuum, expressing the fact that the macroscopic backgrounds are sustained by the quantum fluctuations encoded in $\Gamma$.

\section{Microscopic completion and renormalization group flow}
\subsection{Microscopic higher-derivative gravity with fakeons}
The previous section treated the effective action in a derivative expansion, suitable for energies well below a microscopic cutoff. To construct a UV-complete theory, a microscopic action is needed whose quantization is consistent and renormalizable at arbitrarily high energies. A natural candidate is higher-derivative gravity of the form \cite{Stelle1977,Buchbinder1992}
\begin{align}
S_{\rm grav,micro} &= \int d^4x\sqrt{-g}\Bigl[\frac{M_{\rm Pl}^2}{2}R + \frac{1}{2 f_2^2}C_{\mu\nu\rho\sigma}C^{\mu\nu\rho\sigma} \nonumber\\
&\quad + \frac{1}{3 f_0^2}R^2 + \gamma R_{\mu\nu}R^{\mu\nu} + \cdots\Bigr], \label{eq:S_grav_micro}
\end{align}
where $M_{\rm Pl}$ is the Planck mass, $f_2$ and $f_0$ are dimensionless couplings, and $\gamma$ parametrizes possible additional invariant combinations such as $R_{\mu\nu}R^{\mu\nu}$. Ellipses denote higher-order terms in curvature and its derivatives that may be present in a more general polynomial action.

At the quadratic level around flat space, $g_{\mu\nu}=\eta_{\mu\nu}+\kappa h_{\mu\nu}$ with $\kappa^2=32\pi G$, the propagator for the metric fluctuations can be decomposed into spin-2 and spin-0 parts using projection operators $P^{(2)}$ and $P^{(0-s)}$ \cite{Stelle1977}. In momentum space and in a convenient gauge, the tree-level propagator takes the schematic form
\begin{align}
\tilde{D}_{\mu\nu\rho\sigma}(p) &= \frac{i}{p^2}\Bigl(P^{(2)}_{\mu\nu\rho\sigma} - \tfrac{1}{2}P^{(0-s)}_{\mu\nu\rho\sigma}\Bigr) \nonumber\\
&\quad - \frac{i}{p^2 - m_2^2}P^{(2)}_{\mu\nu\rho\sigma} + \frac{i}{p^2 - m_0^2}\frac{1}{6}P^{(0-s)}_{\mu\nu\rho\sigma} \nonumber\\
&\quad + \text{gauge-dependent terms}, \label{eq:grav_propagator}
\end{align}
where
\begin{align}
m_2^2 &= \frac{M_{\rm Pl}^2}{2 f_2^2},\qquad
m_0^2 = \frac{M_{\rm Pl}^2}{6 f_0^2}. \label{eq:masses_micro}
\end{align}
The poles at $p^2=0$ correspond to the massless graviton, while the poles at $p^2=m_2^2$ and $p^2=m_0^2$ correspond to a massive spin-2 and a massive scalar mode. In the traditional quantization, the residue of the spin-2 pole has the wrong sign (ghost), which jeopardizes unitarity.

The fakeon prescription provides a way to quantize the extra degrees of freedom so that they do not appear as asymptotic states and do not violate unitarity, while maintaining locality and renormalizability at the level of the fundamental action \cite{AnselmiPiva2018}. In this approach, the poles corresponding to the additional degrees of freedom are treated with a specific nonanalytic Wick rotation and average continuation, leading to a unitary S-matrix for the physical states. The resulting theory is perturbatively renormalizable and unitary, with the gravitational couplings $f_2$ and $f_0$ running under the RG in a controlled way.

The matter sector is taken to be the SM minimally coupled to gravity,
\begin{align}
S_{\rm SM,micro} &= \int d^4x\sqrt{-g}\Bigl[\mathcal{L}_{\rm gauge} + \mathcal{L}_{\rm fermion} + \mathcal{L}_{\rm Higgs} + \mathcal{L}_{\rm Yukawa}\Bigr], \label{eq:SSM_micro}
\end{align}
with the usual explicit forms \cite{Glashow1961,Weinberg1967,Salam1968}. Nonminimal couplings of the Higgs field to curvature, such as $\xi_H R |\Phi|^2$, are included in $\mathcal{L}_{\rm Higgs}$. Additional gauge-singlet fermions can be added to account for neutrino masses and dark matter, if desired, without spoiling renormalizability.

The total microscopic action reads
\begin{align}
S_{\rm micro} &= S_{\rm grav,micro} + S_{\rm SM,micro} + S_{\rm gf} + S_{\rm ghost}, \label{eq:Smicro_total}
\end{align}
where $S_{\rm gf}$ and $S_{\rm ghost}$ implement the background field method in the higher-derivative theory. Quantizing $S_{\rm micro}$ with the fakeon prescription defines a UV-complete quantum field theory of gravity and matter. The low-energy effective action $\Gamma$ obtained by integrating out fluctuations in this theory can be matched to the effective action (\ref{eq:Gamma_local}) used in the previous section by integrating out the massive fakeon modes and running the couplings from the microscopic scale down to the desired renormalization scale \cite{AppelquistCarazzone1975}.

\subsection{Functional renormalization group and fixed points}
The functional renormalization group (FRG) provides a nonperturbative tool to explore the RG flow of the effective action in theory space. The central object is the effective average action $\Gamma_k[\bar{g},\bar{A},v_H;h,a,\sigma,\psi]$, which interpolates between the microscopic action at a UV scale $k=\Lambda_{\rm UV}$ and the full effective action at $k\to 0$ \cite{Wetterich1993,Reuter1998,Codello2008,Percacci2017}. The scale $k$ plays the role of an IR cutoff: modes with momenta $p^2\lesssim k^2$ are suppressed by a regulator term, while modes with $p^2\gtrsim k^2$ are integrated out.

The evolution of $\Gamma_k$ with $k$ is governed by the Wetterich equation \cite{Wetterich1993},
\begin{align}
\partial_t \Gamma_k &= \frac{1}{2}\mathrm{STr}\Bigl[\bigl(\Gamma_k^{(2)} + \mathcal{R}_k\bigr)^{-1}\partial_t \mathcal{R}_k\Bigr],\quad t=\ln\frac{k}{k_0}, \label{eq:Wetterich}
\end{align}
where $\Gamma_k^{(2)}$ denotes the Hessian of $\Gamma_k$ with respect to all fluctuating fields, $\mathcal{R}_k$ is an IR regulator, and STr denotes a supertrace over all fields and internal indices with appropriate signs for fermions. The dependence on the backgrounds $\bar{g}$, $\bar{A}$, and $v_H$ is implicit.

To extract RG flows of couplings, a truncation of theory space is chosen. For instance, a simple yet instructive truncation of the gravity--matter system is
\begin{align}
\Gamma_k &= \int d^4x\sqrt{-\bar{g}}\Bigl[\frac{1}{16\pi G_k}\bigl(-\bar{R} + 2\Lambda_k\bigr) \nonumber\\
&\quad + \alpha_{C,k} C_{\mu\nu\rho\sigma}C^{\mu\nu\rho\sigma} + \alpha_{R,k}\bar{R}^2 \nonumber\\
&\quad + \mathcal{L}_{{\rm SM},k}(\bar{g},\bar{A},v_H) + \cdots\Bigr], \label{eq:Gamma_k_trunc}
\end{align}
where $\mathcal{L}_{{\rm SM},k}$ includes the running gauge, Yukawa, and Higgs couplings, and ellipses denote higher-order operators.

Dimensionless couplings are defined by scaling out appropriate powers of $k$, for example
\begin{align}
g_k &= k^2 G_k,\qquad \lambda_k = \frac{\Lambda_k}{k^2}, \nonumber\\
\tilde{\alpha}_{C,k} &= \alpha_{C,k},\qquad \tilde{\alpha}_{R,k} = \alpha_{R,k}. \label{eq:dimless_grav}
\end{align}
Similarly, matter couplings $g_i$, Yukawas $y_a$, and quartic couplings $\lambda_H$ are already dimensionless in four dimensions. Plugging (\ref{eq:Gamma_k_trunc}) into (\ref{eq:Wetterich}) and projecting onto the invariant operators yields a closed system of $\beta$ functions,
\begin{align}
\partial_t g_k &= \beta_g(g_k,\lambda_k,\tilde{\alpha}_{C,k},\tilde{\alpha}_{R,k},\ldots), \nonumber\\
\partial_t \lambda_k &= \beta_\lambda(g_k,\lambda_k,\tilde{\alpha}_{C,k},\tilde{\alpha}_{R,k},\ldots), \nonumber\\
\partial_t \tilde{\alpha}_{C,k} &= \beta_{\alpha_C}(g_k,\lambda_k,\tilde{\alpha}_{C,k},\tilde{\alpha}_{R,k},\ldots), \nonumber\\
\partial_t \tilde{\alpha}_{R,k} &= \beta_{\alpha_R}(g_k,\lambda_k,\tilde{\alpha}_{C,k},\tilde{\alpha}_{R,k},\ldots), \nonumber\\
\partial_t g_i &= \beta_{g_i}(g_i,g_k,\lambda_k,\ldots), \nonumber\\
\partial_t y_a &= \beta_{y_a}(y_a,g_i,g_k,\lambda_k,\ldots), \nonumber\\
\partial_t \lambda_H &= \beta_{\lambda_H}(\lambda_H,y_a,g_i,g_k,\lambda_k,\ldots). \label{eq:beta_system}
\end{align}
The precise functional form of the $\beta$ functions depends on the truncation and the choice of regulator, but qualitative features such as fixed points and the number of relevant directions are robust against reasonable changes \cite{Reuter1998,Codello2008,Percacci2017,Christiansen2018,Eichhorn2019,PastorGutierrez2023,Eichhorn2022Status}.

A fixed point of the RG flow is a point $(g_*,\lambda_*,\tilde{\alpha}_{C,*},\tilde{\alpha}_{R,*},\ldots)$ where all $\beta$ functions vanish. Linearizing the flow around a fixed point yields
\begin{align}
\partial_t \delta u_i &= \sum_j M_{ij}\delta u_j,\quad
M_{ij} = \left.\frac{\partial \beta_{u_i}}{\partial u_j}\right|_{u=u_*}, \label{eq:stability_matrix}
\end{align}
where $u_i$ collectively denotes the couplings. Eigenvalues $\theta_I$ of $-M$ with positive real parts correspond to relevant directions; those with negative real parts correspond to irrelevant directions. The UV critical surface is the subspace spanned by the relevant directions. Trajectories emanating from the UV fixed point and lying within the critical surface are determined by a finite number of parameters corresponding to the projections onto the relevant directions. This is the sense in which an asymptotically safe theory remains predictive despite the presence of infinitely many couplings in principle.

Evidence from FRG computations suggests that gravity with matter admits a nontrivial fixed point with a finite number of relevant directions \cite{Reuter1998,Codello2008,Percacci2017,Christiansen2018,Eichhorn2019,PastorGutierrez2023,Eichhorn2022Status}. In particular, the inclusion of the full SM matter content does not destroy the gravitational fixed point, and the flows of gauge and Yukawa couplings can be influenced by gravitational corrections in such a way that long-standing issues such as the triviality of the hypercharge sector may be resolved \cite{ShaposhnikovWetterich2010,Eichhorn2019,Held2020}.

\subsection{Self-consistency along the RG flow}
The self-consistency postulate can be extended to the scale-dependent effective average action. At each scale $k$, the background fields $\bar{g}_k$, $\bar{A}_k$, and $v_{H,k}$ are defined as the solutions of the scale-dependent gap equations
\begin{align}
\frac{\delta \Gamma_k}{\delta \bar{g}^{\mu\nu}}\Big|_{\bar{g}_k,\bar{A}_k,v_{H,k}} &= 0,\nonumber\\
\frac{\delta \Gamma_k}{\delta \bar{A}_\mu^a}\Big|_{\bar{g}_k,\bar{A}_k,v_{H,k}} &= 0,\nonumber\\
\frac{\delta \Gamma_k}{\delta v_H}\Big|_{\bar{g}_k,\bar{A}_k,v_{H,k}} &= 0. \label{eq:gap_k}
\end{align}
These equations are the scale-dependent analogues of (\ref{eq:gap_vacuum}). They define a trajectory in the space of backgrounds as $k$ is lowered from the UV to the IR. Consistency requires that this trajectory approaches a $k$-independent background as $k\to 0$, reproducing the physical vacuum.

In practice, one often works in a truncation where the backgrounds are restricted to homogeneous and isotropic configurations, such as constant curvature metrics and constant Higgs condensate. In such settings, the gap equations reduce to algebraic equations for a finite number of parameters, such as the curvature scalar $\bar{R}_k$ and the vacuum expectation value $v_{H,k}$. The solvable $O(N)$ model discussed in the next section provides an explicit example. In more general situations, the gap equations become functional equations for the background fields, but the conceptual structure is the same.

The interplay between the RG flow (\ref{eq:beta_system}) and the gap equations (\ref{eq:gap_k}) imposes nontrivial constraints on the couplings. A trajectory in theory space that emanates from the UV fixed point and flows into the IR must be compatible with the existence of a self-consistent vacuum at all intermediate scales. This reduces the number of free parameters compared to a generic RG trajectory and leads to correlations between observables at different scales.

\section{A solvable model: $O(N)$ scalar on constant curvature}
To display the mechanism in a fully analytic setting, consider $N$ real scalars $\phi^I$ with $O(N)$ symmetry and classical potential $V(\phi)=\tfrac{\lambda}{4}(\phi^2 - v_0^2)^2$, nonminimally coupled with coupling $\xi$ to the curvature, on a background of constant scalar curvature $\bar{R}$. We focus on homogeneous backgrounds, so that gradients of the scalar fields vanish.

\subsection{Classical potential and curvature coupling}
Introduce a decomposition
\begin{align}
\phi^I &= \bigl(v + \sigma, \pi^a\bigr),\quad a=1,\dots,N-1, \label{eq:phi_decomp}
\end{align}
where $v$ is the classical expectation value of one component, $\sigma$ is its fluctuation, and $\pi^a$ are the Goldstone modes. The classical Euclidean action relevant for the effective potential is
\begin{align}
S_{\rm cl} &= \int d^4x \sqrt{\bar{g}}\Bigl[\frac{M_{\rm Pl}^2}{2}\bar{R} + \alpha_R \bar{R}^2 - \frac{1}{2}\xi \bar{R}\,\phi^2 \nonumber\\
&\quad + \frac{1}{2}(\bar{\nabla}\phi)^2 + \frac{\lambda}{4}(\phi^2 - v_0^2)^2 + \rho_\Lambda\Bigr], \label{eq:ON_action}
\end{align}
where $\rho_\Lambda$ collects the cosmological constant term and constant counterterms. In homogeneous backgrounds, $\bar{\nabla}\phi=0$, so only the potential terms contribute. Substituting $\phi^2=v^2$ in this configuration, the tree-level potential density is
\begin{align}
U_{\rm tree}(v,\bar{R}) &= -\frac{M_{\rm Pl}^2}{2}\bar{R} - \alpha_R \bar{R}^2 + \frac{1}{2}\xi \bar{R}\,v^2 \nonumber\\
&\quad + \frac{\lambda}{4}(v^2 - v_0^2)^2 + \rho_\Lambda. \label{eq:Utree}
\end{align}
The first two terms arise from the gravitational sector, the third from the nonminimal coupling, and the fourth and fifth from the scalar self-interaction and cosmological constant.

\subsection{One-loop effective potential with curvature dependence}
Quantum fluctuations of $\sigma$ and $\pi^a$ around the homogeneous background $(v,\bar{R})$ generate corrections to the effective potential. To one loop, each fluctuating mode contributes a term of the form \cite{BirrellDavies1982,ParkerToms2009,Buchbinder1992}
\begin{align}
\Delta U &= \frac{1}{2}\int\frac{d^4p}{(2\pi)^4}\ln\bigl(p^2 + m^2(v,\bar{R})\bigr), \label{eq:DeltaU_basic}
\end{align}
where $m^2(v,\bar{R})$ is the field-dependent mass of the mode. Using dimensional regularization and the $\overline{\rm MS}$ scheme, this integral evaluates to
\begin{align}
\Delta U &= \frac{1}{64\pi^2} m^4(v,\bar{R})\Bigl[\ln\frac{m^2(v,\bar{R})}{\mu^2} - \frac{3}{2}\Bigr] \nonumber\\
&\quad + \text{counterterms}, \label{eq:DeltaU_MSbar}
\end{align}
where $\mu$ is the renormalization scale and the counterterms are chosen to absorb the divergences and define the renormalized couplings.

In the presence of curvature, the field-dependent masses of the $\pi^a$ and $\sigma$ modes are
\begin{align}
m_\pi^2(v,\bar{R}) &= \lambda(v^2 - v_0^2) + \xi \bar{R}, \label{eq:mpi}\\
m_\sigma^2(v,\bar{R}) &= \lambda(3v^2 - v_0^2) + \xi \bar{R}. \label{eq:msig}
\end{align}
There are $N-1$ degenerate $\pi^a$ modes and one $\sigma$ mode. Summing their contributions, the renormalized one-loop correction in the $\overline{\rm MS}$ scheme can be written as
\begin{align}
U_{\rm 1\mbox{-}loop}(v,\bar{R}) &= \sum_{i=\pi,\sigma}\frac{n_i}{64\pi^2} m_i^4(v,\bar{R}) \nonumber\\
&\quad \times \Bigl[\ln\frac{m_i^2(v,\bar{R})}{\mu^2} - c_i\Bigr] \nonumber\\
&\quad + U_{\rm ct}(v,\bar{R}), \label{eq:U1loop_full}
\end{align}
with $n_\pi=N-1$, $n_\sigma=1$, and $c_i=\tfrac{3}{2}$ for scalar fields in the $\overline{\rm MS}$ scheme. The counterterm contribution $U_{\rm ct}(v,\bar{R})$ absorbs the UV divergences and can be chosen to enforce convenient renormalization conditions, for example fixing $U_{\rm eff}$ and its derivatives at reference values of $(v,\bar{R})$.

To leading order in curvature and for slowly varying backgrounds, terms involving explicit factors of $\bar{R}$ beyond those in the masses $m_i^2$ can be neglected or absorbed into the running of $\alpha_R$ and $\xi$. Then $U_{\rm 1\mbox{-}loop}$ can be approximated by
\begin{align}
U_{\rm 1\mbox{-}loop}(v,\bar{R}) &\simeq \sum_{i=\pi,\sigma}\frac{n_i}{64\pi^2} m_i^4(v,\bar{R}) \nonumber\\
&\quad \times \Bigl[\ln\frac{m_i^2(v,\bar{R})}{\mu^2} - c_i\Bigr], \label{eq:U1loop}
\end{align}
with the understanding that curvature-squared terms generated by the loop have been absorbed into a redefinition of $\alpha_R(\mu)$. A more complete expression including explicit $\bar{R}$ dependence is summarized in Appendix E.

The full effective potential is then
\begin{align}
U_{\rm eff}(v,\bar{R}) &= U_{\rm tree}(v,\bar{R}) + U_{\rm 1\mbox{-}loop}(v,\bar{R}). \label{eq:Ueff}
\end{align}

\subsection{Gap equations and their explicit form}
The vacuum gap equations for homogeneous backgrounds are the conditions that $U_{\rm eff}$ be stationary with respect to variations of $v$ and $\bar{R}$,
\begin{align}
\frac{\partial U_{\rm eff}}{\partial v} &= 0,\quad
\frac{\partial U_{\rm eff}}{\partial \bar{R}} = 0. \label{eq:gap_vR}
\end{align}
These equations determine $(v,\bar{R})$ in terms of the renormalized couplings $(\lambda,\xi,\alpha_R,M_{\rm Pl}^2,\rho_\Lambda)$ and the scale $\mu$.

To obtain the explicit form of $\partial U_{\rm eff}/\partial v$, differentiate (\ref{eq:Utree}) and (\ref{eq:U1loop}) with respect to $v$. The tree-level contribution is
\begin{align}
\frac{\partial U_{\rm tree}}{\partial v} &= \frac{1}{2}\xi \bar{R}\,(2v) + \frac{\lambda}{4}\cdot 2(v^2 - v_0^2)\cdot 2v \nonumber\\
&= \xi \bar{R}\,v + \lambda(v^2 - v_0^2)v. \label{eq:dUtree_dv}
\end{align}
For the one-loop term, consider a single mode with mass $m^2(v,\bar{R})$. Its contribution to $U_{\rm 1\mbox{-}loop}$ is
\begin{align}
\Delta U &= \frac{n}{64\pi^2}m^4\Bigl[\ln\frac{m^2}{\mu^2} - c\Bigr], \label{eq:DeltaU_single}
\end{align}
where $n$ is the multiplicity and $c$ is a constant. Differentiating with respect to $v$,
\begin{align}
\frac{\partial \Delta U}{\partial v} &= \frac{n}{64\pi^2}\Bigl[4 m^3 \frac{\partial m}{\partial v}\Bigl(\ln\frac{m^2}{\mu^2} - c\Bigr) \nonumber\\
&\quad + m^4\cdot \frac{1}{m^2}\cdot 2m\frac{\partial m}{\partial v}\Bigr]. \label{eq:dDeltaU_dv_step1}
\end{align}
Using $\partial m/\partial v = (\partial m^2/\partial v)/(2m)$, this simplifies to
\begin{align}
\frac{\partial \Delta U}{\partial v} &= \frac{n}{64\pi^2}\Bigl[2 m^2\frac{\partial m^2}{\partial v}\Bigl(\ln\frac{m^2}{\mu^2} - c\Bigr) \nonumber\\
&\quad + m^2\frac{\partial m^2}{\partial v}\Bigr]. \label{eq:dDeltaU_dv_step2}
\end{align}
Choosing $c=\tfrac{3}{2}$ as in the $\overline{\rm MS}$ scheme, one can rewrite
\begin{align}
\ln\frac{m^2}{\mu^2} - c + \frac{1}{2} &= \ln\frac{m^2}{\mu^2} - 1. \label{eq:log_shift}
\end{align}
Thus
\begin{align}
\frac{\partial \Delta U}{\partial v} &= \frac{n}{32\pi^2} m^2\frac{\partial m^2}{\partial v}\Bigl[\ln\frac{m^2}{\mu^2} - 1\Bigr]. \label{eq:dDeltaU_dv_final}
\end{align}
Summing over $i=\pi,\sigma$ with multiplicities $n_i$, one obtains
\begin{align}
\frac{\partial U_{\rm 1\mbox{-}loop}}{\partial v} &= \sum_{i=\pi,\sigma}\frac{n_i}{32\pi^2} m_i^2(v,\bar{R})\frac{\partial m_i^2(v,\bar{R})}{\partial v} \nonumber\\
&\quad \times \Bigl[\ln\frac{m_i^2(v,\bar{R})}{\mu^2} - 1\Bigr]. \label{eq:dU1loop_dv}
\end{align}
Using (\ref{eq:mpi}) and (\ref{eq:msig}),
\begin{align}
\frac{\partial m_\pi^2}{\partial v} &= 2\lambda v,\quad
\frac{\partial m_\sigma^2}{\partial v} = 6\lambda v. \label{eq:dm_dv}
\end{align}
Combining (\ref{eq:dUtree_dv}) and (\ref{eq:dU1loop_dv}), the gap equation $\partial U_{\rm eff}/\partial v=0$ becomes
\begin{align}
0 &= \lambda(v^2 - v_0^2)v + \xi \bar{R}\,v \nonumber\\
&\quad + \sum_{i=\pi,\sigma}\frac{n_i}{32\pi^2} m_i^2(v,\bar{R})\frac{\partial m_i^2(v,\bar{R})}{\partial v} \nonumber\\
&\quad \times \Bigl[\ln\frac{m_i^2(v,\bar{R})}{\mu^2} - 1\Bigr]. \label{eq:gap_v_explicit}
\end{align}
This is equation (8) of the main text written with all intermediate steps displayed.

Next, consider $\partial U_{\rm eff}/\partial \bar{R}$. The tree-level contribution from (\ref{eq:Utree}) is
\begin{align}
\frac{\partial U_{\rm tree}}{\partial \bar{R}} &= -\frac{M_{\rm Pl}^2}{2} - 2\alpha_R \bar{R} + \frac{1}{2}\xi v^2. \label{eq:dUtree_dR}
\end{align}
For a single mode, differentiating (\ref{eq:DeltaU_single}) with respect to $\bar{R}$ yields
\begin{align}
\frac{\partial \Delta U}{\partial \bar{R}} &= \frac{n}{64\pi^2}\Bigl[4 m^3 \frac{\partial m}{\partial \bar{R}}\Bigl(\ln\frac{m^2}{\mu^2} - c\Bigr) \nonumber\\
&\quad + m^4\cdot \frac{1}{m^2}\cdot 2m\frac{\partial m}{\partial \bar{R}}\Bigr]. \label{eq:dDeltaU_dR_step1}
\end{align}
Proceeding as before and using $\partial m/\partial \bar{R} = (\partial m^2/\partial \bar{R})/(2m)$, one finds
\begin{align}
\frac{\partial \Delta U}{\partial \bar{R}} &= \frac{n}{32\pi^2} m^2\frac{\partial m^2}{\partial \bar{R}}\Bigl[\ln\frac{m^2}{\mu^2} - 1\Bigr]. \label{eq:dDeltaU_dR_final}
\end{align}
For the $O(N)$ model, $\partial m_i^2/\partial \bar{R}=\xi$ for both $i=\pi,\sigma$. Summing over modes,
\begin{align}
\frac{\partial U_{\rm 1\mbox{-}loop}}{\partial \bar{R}} &= \sum_{i=\pi,\sigma}\frac{n_i}{32\pi^2} m_i^2(v,\bar{R})\xi\Bigl[\ln\frac{m_i^2(v,\bar{R})}{\mu^2} - 1\Bigr]. \label{eq:dU1loop_dR}
\end{align}
Combining (\ref{eq:dUtree_dR}) and (\ref{eq:dU1loop_dR}), the gap equation $\partial U_{\rm eff}/\partial \bar{R}=0$ becomes
\begin{align}
0 &= -\frac{M_{\rm Pl}^2}{2} - 2\alpha_R \bar{R} + \frac{\xi}{2} v^2 \nonumber\\
&\quad + \sum_{i=\pi,\sigma}\frac{n_i}{32\pi^2} m_i^2(v,\bar{R})\xi\Bigl[\ln\frac{m_i^2(v,\bar{R})}{\mu^2} - 1\Bigr]. \label{eq:gap_R_explicit}
\end{align}
Equations (\ref{eq:gap_v_explicit}) and (\ref{eq:gap_R_explicit}) together determine $(v,\bar{R})$ as functions of $(\lambda,\xi,\alpha_R,M_{\rm Pl}^2,\rho_\Lambda,\mu)$.

\subsection{Limiting cases and de Sitter branch}
Two limiting cases illustrate the physics encoded in the gap equations.

First, consider the Minkowski vacuum $\bar{R}=0$. Neglecting the loop terms at a scale $\mu\sim v$ where they are small, (\ref{eq:gap_v_explicit}) reduces to
\begin{align}
\lambda(v^2 - v_0^2)v &\simeq 0. \label{eq:gap_v_Mink}
\end{align}
Besides the trivial solution $v=0$, the symmetry-breaking solution $v\simeq v_0$ is selected by phenomenology. In this branch, (\ref{eq:gap_R_explicit}) with $\bar{R}=0$ becomes
\begin{align}
0 &= -\frac{M_{\rm Pl}^2}{2} + \frac{\xi}{2} v_0^2 + \delta \rho_\Lambda, \label{eq:gap_R_Mink}
\end{align}
where $\delta \rho_\Lambda$ represents the sum of the loop contribution and the finite part of the cosmological constant counterterm. This equation fixes the renormalized cosmological constant to ensure a flat vacuum, reproducing the familiar tuning of the vacuum energy.

Second, consider a de Sitter branch with $\bar{R}>0$. Neglect the loop contributions and assume $v\neq 0$. Equation (\ref{eq:gap_v_explicit}) reduces to
\begin{align}
\lambda(v^2 - v_0^2) + \xi \bar{R} &= 0. \label{eq:gap_v_dS_tree}
\end{align}
Equation (\ref{eq:gap_R_explicit}) becomes
\begin{align}
-\frac{M_{\rm Pl}^2}{2} - 2\alpha_R \bar{R} + \frac{\xi}{2} v^2 &= 0. \label{eq:gap_R_dS_tree}
\end{align}
Solving (\ref{eq:gap_v_dS_tree}) for $v^2$ and substituting into (\ref{eq:gap_R_dS_tree}) yields
\begin{align}
-\frac{M_{\rm Pl}^2}{2} - 2\alpha_R \bar{R} + \frac{\xi}{2}\Bigl(v_0^2 - \frac{\xi}{\lambda}\bar{R}\Bigr) &= 0. \label{eq:gap_R_dS_sub}
\end{align}
Rearranging,
\begin{align}
\Bigl(2\alpha_R + \frac{\xi^2}{2\lambda}\Bigr)\bar{R} &= \frac{\xi}{2}v_0^2 - \frac{M_{\rm Pl}^2}{2}. \label{eq:R_dS_solution}
\end{align}
Assuming the denominator is positive,
\begin{align}
\bar{R} &= \frac{\xi v_0^2 - M_{\rm Pl}^2}{4\alpha_R + \xi^2/\lambda}. \label{eq:R_dS_explicit}
\end{align}
For suitable choices of parameters, this solution yields a positive curvature de Sitter vacuum. Loop corrections shift the relation between $v$ and $\bar{R}$ but do not qualitatively alter the existence of such a branch. This de Sitter branch underlies the $R^2$ inflation scenario discussed in the next section, where $\alpha_R$ is large and positive and curvature is dominated by the $R^2$ term.

The $O(N)$ model thus provides a concrete illustration of how the self-consistency postulate translates into coupled gap equations that fix both the condensate and the curvature in terms of renormalized couplings. The same logic extends to the full SM plus gravity, with the technical difference that gauge bosons and fermions contribute additional terms and thresholds.

\section{Phenomenology and predictions}
\subsection{Inflation from the induced $R^2$ term}
The universal generation of curvature-squared operators is unavoidable at one loop. The $R^2$ term is particularly important because, when dominant, it is dynamically equivalent to GR plus a scalar field (the scalaron) and is ghost free in the scalar sector \cite{Starobinsky1980,Whitt1984,Stelle1977}. In the Jordan frame, consider
\begin{align}
S_{R^2} &= \int d^4x\sqrt{-g}\Bigl[\frac{M_{\rm Pl}^2}{2}R + \alpha_R R^2\Bigr]. \label{eq:SR2}
\end{align}
Introducing an auxiliary field $\chi$,
\begin{align}
S_{R^2} &= \int d^4x\sqrt{-g}\Bigl[\frac{M_{\rm Pl}^2}{2}R + 2\alpha_R\chi R - \alpha_R\chi^2\Bigr], \label{eq:SR2_aux}
\end{align}
and performing a Weyl rescaling to the Einstein frame,
\begin{align}
g_{\mu\nu} &\to \tilde{g}_{\mu\nu} = \Omega^2 g_{\mu\nu},\quad
\Omega^2 = 1 + \frac{4\alpha_R \chi}{M_{\rm Pl}^2}, \label{eq:Weyl_R2}
\end{align}
brings the action to the Einstein form plus a scalar with canonical kinetic term. Defining the scalaron field $\varphi$ by
\begin{align}
\Omega^2 &= e^{\sqrt{\frac{2}{3}}\varphi/M_{\rm Pl}}, \label{eq:varphi_def}
\end{align}
one finds in the Einstein frame the potential \cite{Whitt1984}
\begin{align}
V(\varphi) &= \frac{3}{4} M^2 M_{\rm Pl}^2 \Bigl(1 - e^{-\sqrt{\frac{2}{3}}\frac{\varphi}{M_{\rm Pl}}}\Bigr)^2, \label{eq:Starobinsky_potential}
\end{align}
where
\begin{align}
\alpha_R &= \frac{M_{\rm Pl}^2}{12 M^2}. \label{eq:alpha_relation}
\end{align}
During inflation, the field $\varphi$ slowly rolls on the plateau of the potential at large positive values. The slow-roll parameters computed from (\ref{eq:Starobinsky_potential}) yield, for $N_e$ e-folds of inflation,
\begin{align}
n_s &\simeq 1 - \frac{2}{N_e}, \qquad r \simeq \frac{12}{N_e^2}. \label{eq:ns_r}
\end{align}
For $N_e=50$--$60$, this gives $n_s\simeq 0.962$--$0.967$ and $r\simeq 0.003$--$0.004$, well within the Planck constraints \cite{Planck2018}. The amplitude of scalar perturbations fixes the mass scale $M$ to be
\begin{align}
M &\simeq (1.3 \pm 0.1)\times 10^{-5} M_{\rm Pl}, \label{eq:M_value}
\end{align}
and hence
\begin{align}
\alpha_R &\simeq (4.5\text{--}5.5)\times 10^8. \label{eq:alphaR_value}
\end{align}

In the present framework, $\alpha_R$ is not an arbitrary parameter but is generated and renormalized by quantum fluctuations of SM fields and gravity. The value of $\alpha_R$ at inflationary scales is determined by running from the UV fixed point down to those scales and by matching across particle thresholds. While a full computation requires solving the RG equations including gravitational contributions \cite{ShaposhnikovWetterich2010,EichhornHeldPauly2021,PastorGutierrez2023}, the qualitative point is that the coefficient required by inflation is compatible with natural values of the curvature-squared couplings at high scales in asymptotically safe gravity--matter systems \cite{Eichhorn2019,Held2020}. The inflationary sector of the theory is therefore sharply predictive and does not require new degrees of freedom beyond those already present in the higher-derivative gravitational action.

\subsection{Short-distance gravity and laboratory bounds}
Quadratic gravity around flat space contains, besides the massless graviton, a massive scalar (the scalaron) and a massive spin-2 mode from the $C^2$ term \cite{Stelle1977}. At tree level, the static potential between two point masses $m_1$ and $m_2$ separated by distance $r$ in the nonrelativistic limit has the form \cite{Stelle1977,Donoghue1994}
\begin{align}
V(r) &= -\frac{G_N m_1 m_2}{r}\Biggl[1 + \frac{1}{3}e^{-M r} - \frac{4}{3}e^{-m_2 r} \nonumber\\
&\quad + \frac{41}{10\pi}\frac{G_N \hbar}{r^2 c^3} + \cdots \Biggr]. \label{eq:potential}
\end{align}
The first term is the Newtonian potential, the second and third are Yukawa corrections from the scalar and spin-2 modes, and the last term is the universal quantum correction computed in effective field theory \cite{Donoghue1994}.

Submillimeter tests of Newton's law constrain Yukawa corrections to have ranges $\lambda \lesssim 0.1$ mm for order-one strength \cite{Kapner2007}. This implies $M, m_2 \gtrsim 2\times 10^{-3}$ eV. In the inflationary regime, the scalaron mass is $M\sim 10^{13}$ GeV according to (\ref{eq:M_value}), and typical loop-induced values of $\alpha_C$ correspond to $m_2\sim M_{\rm Pl}/\sqrt{\alpha_C}$ well above laboratory energies. Thus the Yukawa corrections to Newton's law are completely negligible at accessible distances and the theory is consistent with all existing tests of the inverse-square law.

The quantum correction proportional to $1/r^3$ in (\ref{eq:potential}) is a universal prediction of low-energy quantum gravity, independent of the UV completion \cite{Donoghue1994}. Its smallness makes it unobservable in current experiments, but it is a characteristic signature of the quantum vacuum self-consistency framework, tying together the microscopic theory and its long-distance effective behavior.

\subsection{Gravitational waves}
In the infrared limit where higher-derivative contributions to the field equations are suppressed, gravitational waves propagate at the speed of light on cosmological and astrophysical backgrounds. This is in agreement with the multimessenger observation of GW170817 and its electromagnetic counterpart, which constrain the difference between the speeds of gravity and light to be extremely small at frequencies around $10^2$ Hz \cite{GW170817}. In the present framework, the dispersion relations of the massive scalar and spin-2 modes are modified by their large masses $M$ and $m_2$, so that they are not excited at LIGO/Virgo/KAGRA frequencies. Nonlocal form factors induce a mild running of $G_N$ with momentum but do not alter the luminal propagation of the massless graviton in the observed regime \cite{Donoghue1994,BarvinskyVilkovisky1990}.

At much higher frequencies, close to the masses of the scalaron and spin-2 modes, new phenomena such as the excitation of these modes and their subsequent decay could occur. However, such frequencies are far beyond current experimental reach. In any case, the fakeon quantization prescription ensures that the extra modes do not appear as asymptotic states and do not compromise unitarity \cite{AnselmiPiva2018}.

\subsection{Running couplings and vacuum stability}
The RG flow of SM couplings up to the Planck scale has been studied extensively \cite{Buttazzo2013}. In the absence of gravity, the Higgs quartic $\lambda_H$ runs to small or even negative values around $10^{9}$--$10^{11}$ GeV for the measured Higgs and top masses, placing the electroweak vacuum near the boundary of metastability. The inclusion of gravitational corrections modifies the $\beta$ functions of the gauge, Yukawa, and scalar couplings above a scale of order $M_{\rm Pl}$ \cite{ShaposhnikovWetterich2010,EichhornHeldPauly2021}. Schematically,
\begin{align}
\mu\frac{d x_j}{d\mu} &= \beta_j^{\rm SM}(x) + \beta_j^{\rm grav}(x,g,\lambda,\ldots), \label{eq:beta_grav_matter}
\end{align}
where $x_j$ collectively denotes SM couplings and $\beta_j^{\rm grav}$ encodes gravitational contributions, whose detailed form depends on the RG scheme and truncation.

If the combined gravity--matter system has an asymptotically safe fixed point, the high-energy behavior of $\lambda_H$ can be controlled by that fixed point. For example, in scenarios where the gravity-induced anomalous dimension of $\lambda_H$ is positive, the RG flow can drive $\lambda_H$ to zero at the Planck scale, yielding a prediction for the Higgs mass in the vicinity of the vacuum stability bound \cite{ShaposhnikovWetterich2010,EichhornHeldPauly2021}. The self-consistency gap equations then require that the Higgs condensate and curvature adjust so that the vacuum lies at the minimum of the effective potential, providing additional constraints on the allowed values of the couplings.

A complete determination of the RG flow including gravity and the imposition of the gap equations across scales is technically demanding and remains an active area of research \cite{Eichhorn2019,PastorGutierrez2023,Eichhorn2022Status}. Nevertheless, the qualitative picture is clear: the self-consistency principle combined with asymptotic safety reduces the number of free parameters and correlates observables across sectors.

\subsection{Uniqueness and distinctive predictions}
The framework developed here has several distinctive features that set it apart from other approaches to quantum gravity and unification:
\begin{itemize}
\item Classical backgrounds of gravity and the SM are unified as macroscopic order parameters of a single quantum vacuum. The self-consistency postulate requires that these backgrounds be solutions of gap equations derived from the effective action. This leads to relations between the curvature, the Higgs vacuum expectation value, and the values of couplings at the vacuum.

\item The gravitational sector is UV complete without introducing new massless degrees of freedom beyond the metric. The higher-derivative terms required by renormalizability are present, but their extra degrees of freedom are quantized as fakeons and do not appear as physical particles. This differs from string theory and other approaches where an infinite tower of new states appears.

\item The existence of an asymptotically safe fixed point in the combined gravity--matter system reduces the number of free parameters to a finite set of relevant couplings. Together with the gap equations, this yields correlations among low-energy observables such as the Higgs and top masses, the inflationary parameters $(n_s,r)$, and the coefficients of higher-derivative gravitational operators.

\item The inflationary sector is fully determined by the induced $R^2$ term, leading to the Starobinsky predictions (\ref{eq:ns_r}) and (\ref{eq:M_value}) without free parameters specific to inflation. This is more predictive than many models of inflation that introduce ad hoc scalar fields and potentials.

\item The universal quantum correction to Newton's potential and the pattern of Yukawa corrections from massive gravitational modes are fixed once the microscopic couplings are specified. These corrections provide potential targets for future high-precision tests of gravity, although they are currently beyond reach.

\end{itemize}
These features make the quantum vacuum self-consistency framework a concrete and testable candidate for a theory of all fundamental interactions, while maintaining a conservative quantum field theoretic foundation.

\section{Predictivity, renormalization conditions, and limitations}
\subsection{Renormalization conditions and vacuum selection}
Predictivity arises in several ways:
\begin{itemize}
\item The $R^2$-driven inflationary sector has fixed $(n_s,r)$ once the number of e-folds $N_e$ is specified and the amplitude of scalar perturbations is matched, leading to the determination of $M$ and $\alpha_R$.

\item The universal quantum correction to Newton's potential has a fixed coefficient independent of microscopic details.

\item The strengths and ranges of Yukawa corrections to gravity from the scalaron and spin-2 modes are determined by the masses $M$ and $m_2$, which in turn are functions of the microscopic couplings $f_0$ and $f_2$.

\item The existence of a UV fixed point and the finite number of relevant directions constrain the high-energy values of SM couplings and, through RG flow, their low-energy values.
\end{itemize}
Within the full coupled system, renormalization conditions encode the choice of vacuum. A natural set of conditions for homogeneous backgrounds is
\begin{align}
\frac{\partial U_{\rm eff}}{\partial v}\Big|_{v=v_\star,\bar{R}=\bar{R}_\star} &= 0,\nonumber\\
\frac{\partial U_{\rm eff}}{\partial \bar{R}}\Big|_{v=v_\star,\bar{R}=\bar{R}_\star} &= 0,\nonumber\\
U_{\rm eff}(v_\star,\bar{R}_\star) &= \rho_{\rm vac}^{\rm obs}, \label{eq:renorm_conditions}
\end{align}
where $(v_\star,\bar{R}_\star)$ describe the observed vacuum and $\rho_{\rm vac}^{\rm obs}$ is the observed vacuum energy density. The first two equations are the vacuum gap equations, while the third fixes the finite part of the cosmological constant counterterm.

In practical applications, one matches the renormalized couplings to measured low-energy observables at a reference scale (for example, the $Z$ mass) and then evolves them upward using the RG equations, imposing (\ref{eq:renorm_conditions}) at a vacuum scale (for example, the electroweak or inflationary scale). Conversely, under the assumption of a UV fixed point, one can start at the fixed point, follow the RG flow down, and see whether a vacuum satisfying (\ref{eq:renorm_conditions}) exists. This procedure can rule out large regions of parameter space and in favorable cases can lead to sharp predictions.

\subsection{Limitations and open issues}
Several important limitations and open issues remain:
\begin{itemize}
\item The existence and detailed structure of the asymptotically safe fixed point in the full gravity--SM system are supported by FRG computations in truncated theory spaces but are not yet rigorously established. Different truncations and regulators can lead to quantitatively different flows, and convergence of the truncation must be carefully assessed \cite{Eichhorn2019,Eichhorn2022Status,Bonanno2020}.

\item Quadratic gravity is perturbatively renormalizable but, in the traditional quantization, contains a massive spin-2 ghost. The fakeon prescription provides a proposal to remove this ghost from the physical spectrum while preserving unitarity and locality at the level of the microscopic action \cite{AnselmiPiva2018}. Its full nonperturbative status remains to be understood.

\item The smallness of the observed cosmological constant is not explained within the present framework. The condition $U_{\rm eff}(v_\star,\bar{R}_\star)=\rho_{\rm vac}^{\rm obs}$ requires a delicate cancellation between large contributions from vacuum fluctuations and counterterms, as in other semiclassical gravity approaches.

\item Threshold matching in curved space, nonlocal form factors, and finite-temperature effects in the early universe require careful treatment. While the covariant perturbation theory of Barvinsky and Vilkovisky provides the conceptual basis, explicit computations in the full SM plus gravity are technically involved.

\item The impact of asymptotically safe gravity on SM couplings, including the Higgs quartic and top Yukawa, depends on details of the gravitational sector and the truncation. Predictions for quantities such as the Higgs mass require further refinement and cross-checking \cite{ShaposhnikovWetterich2010,EichhornHeldPauly2021,Eichhorn2019,PastorGutierrez2023}.
\end{itemize}
Despite these caveats, the quantum vacuum self-consistency principle embedded in a UV-complete gravity--matter theory yields a cohesive and predictive picture that connects inflationary cosmology, precision gravity, and particle physics within a single calculational framework.

\section{Conclusion}
Starting from the postulate that the classical backgrounds of gravity and the Standard Model are macroscopic order parameters of a unified quantum vacuum sustained by its own fluctuations, a coherent, calculable, and ultraviolet-complete framework has been constructed. The renormalized Einstein, Yang--Mills, and Higgs equations emerge as vacuum equations of state derived from the stationarity of the one-particle-irreducible effective action. The heat-kernel expansion fixes the renormalization of all couplings and requires the presence of higher-derivative gravitational terms. A solvable $O(N)$ model illustrates how the vacuum gap equations determine both curvature and condensates.

Embedding the framework into higher-derivative gravity quantized with the fakeon prescription and coupled to the Standard Model provides a perturbatively renormalizable and unitary microscopic theory. Functional renormalization group techniques support the existence of an interacting UV fixed point for the combined gravity--matter system, rendering the theory asymptotically safe and predictive. The theory makes concrete predictions, notably Starobinsky-type inflation with fixed $(n_s,r)$, universal quantum corrections to gravity, and correlations between high-energy and low-energy observables in the particle physics sector. It is consistent with current astrophysical and laboratory constraints and reduces to General Relativity and the Standard Model at accessible scales with controlled corrections. A systematic program of higher-loop computations, improved FRG truncations, and phenomenological applications can further test and refine the picture.

\appendix

\section{Conventions and useful variations}
We use signature $(-,+,+,+)$ and define the Riemann tensor, Ricci tensor, and scalar curvature as
\begin{align}
R^\rho{}_{\sigma\mu\nu} &= \partial_\mu \Gamma^\rho_{\nu\sigma} - \partial_\nu \Gamma^\rho_{\mu\sigma} + \Gamma^\rho_{\mu\lambda}\Gamma^\lambda_{\nu\sigma} - \Gamma^\rho_{\nu\lambda}\Gamma^\lambda_{\mu\sigma}, \label{eq:Riemann_def}\\
R_{\mu\nu} &= R^\rho{}_{\mu\rho\nu},\qquad
R = \bar{g}^{\mu\nu}R_{\mu\nu}. \label{eq:Ricci_R_def}
\end{align}
The Weyl tensor is
\begin{align}
C_{\mu\nu\rho\sigma} &= R_{\mu\nu\rho\sigma} - \frac{1}{2}\bigl(\bar{g}_{\mu\rho}R_{\nu\sigma} - \bar{g}_{\mu\sigma}R_{\nu\rho} \nonumber\\
&\quad - \bar{g}_{\nu\rho}R_{\mu\sigma} + \bar{g}_{\nu\sigma}R_{\mu\rho}\bigr) \nonumber\\
&\quad + \frac{R}{6}\bigl(\bar{g}_{\mu\rho}\bar{g}_{\nu\sigma} - \bar{g}_{\mu\sigma}\bar{g}_{\nu\rho}\bigr). \label{eq:Weyl_def}
\end{align}
The basic metric variations used in the main text are
\begin{align}
\delta \bigl(\sqrt{-\bar{g}}\,R\bigr) &= \sqrt{-\bar{g}}\Bigl[\bigl(G_{\mu\nu} + \bar{g}_{\mu\nu}\Box - \nabla_\mu\nabla_\nu\bigr) \delta \bar{g}^{\mu\nu}\Bigr], \label{eq:delta_R}\\
\delta \bigl(\sqrt{-\bar{g}}\,R^2\bigr) &= \sqrt{-\bar{g}}\,H^{(R^2)}_{\mu\nu}\delta \bar{g}^{\mu\nu}, \label{eq:delta_R2}\\
\delta \bigl(\sqrt{-\bar{g}}\,C^2\bigr) &= 2\sqrt{-\bar{g}}\,B_{\mu\nu}\delta \bar{g}^{\mu\nu}. \label{eq:delta_C2}
\end{align}
The explicit forms of $H^{(R^2)}_{\mu\nu}$ and $B_{\mu\nu}$ are given in (\ref{eq:HR2}) and (\ref{eq:Bach}).

\section{Heat-kernel coefficients for standard fields}
For a real scalar with nonminimal coupling $\xi$, a Dirac fermion, and a gauge vector (in background gauge), the $a_2$ densities contributing to $C^2$ and $E_4$ are, respectively \cite{Duff1994,Buchbinder1992},
\begin{align}
a_{2}^{\rm scalar} &\supset \frac{1}{120}C^2 - \frac{1}{360}E_4 + \frac{1}{2}\Bigl(\xi-\tfrac{1}{6}\Bigr)^2 R^2, \label{eq:a2_scalar}\\
a_{2}^{\rm Dirac} &\supset \frac{1}{20}C^2 - \frac{11}{360}E_4, \label{eq:a2_Dirac}\\
a_{2}^{\rm vector} &\supset \frac{1}{10}C^2 - \frac{31}{180}E_4. \label{eq:a2_vector}
\end{align}
Gauge and ghost contributions combine to the vector result.

\section{Bach tensor and equations of motion}
The Bach tensor defined in (\ref{eq:Bach}) is traceless, $B^\mu{}_\mu=0$, and vanishes in conformally flat spacetimes. In four dimensions, the metric variation of $\int d^4x\sqrt{-g}\,C^2$ yields $2\int d^4x\sqrt{-g}\,B_{\mu\nu}\delta g^{\mu\nu}$. Combining this with (\ref{eq:delta_R}) and (\ref{eq:delta_R2}) yields the higher-derivative contributions to the field equations (\ref{eq:Einstein_HD}).

\section{Standard Model one-loop beta functions}
At one loop in the $\overline{\rm MS}$ scheme, the SM gauge couplings obey \cite{MachacekVaughnI,MachacekVaughnII}
\begin{align}
\mu\frac{dg_1}{d\mu} &= \frac{41}{6}\frac{g_1^3}{16\pi^2}, \label{eq:beta_g1}\\
\mu\frac{dg_2}{d\mu} &= -\frac{19}{6}\frac{g_2^3}{16\pi^2}, \label{eq:beta_g2}\\
\mu\frac{dg_3}{d\mu} &= -7\frac{g_3^3}{16\pi^2}, \label{eq:beta_g3}
\end{align}
with GUT normalization for $g_1$. The Higgs quartic $\lambda_H$ and the top Yukawa $y_t$ run as \cite{MachacekVaughnI,MachacekVaughnII,Buttazzo2013}
\begin{align}
\mu\frac{d\lambda_H}{d\mu} &= \frac{1}{16\pi^2}\Bigl[24\lambda_H^2 - 6y_t^4 \nonumber\\
&\quad + \frac{3}{8}\bigl(2g_2^4 + (g_2^2 + g_1^2)^2\bigr) \nonumber\\
&\quad - 9\lambda_H g_2^2 - 3\lambda_H g_1^2 + 12\lambda_H y_t^2 \Bigr], \label{eq:beta_lambdaH}\\
\mu\frac{dy_t}{d\mu} &= \frac{y_t}{16\pi^2}\Bigl[\frac{9}{2}y_t^2 - \frac{17}{12}g_1^2 - \frac{9}{4}g_2^2 - 8 g_3^2 \Bigr]. \label{eq:beta_yt}
\end{align}

\section{One-loop effective potential with curvature dependence}
For a scalar field with field-dependent mass $m^2(\phi,R)=m_0^2(\phi)+\xi R$, the one-loop contribution to the effective potential in the limit of slowly varying backgrounds is \cite{BirrellDavies1982,ParkerToms2009,Buchbinder1992}
\begin{align}
\Delta U &= \frac{1}{64\pi^2} m^4(\phi,R)\Bigl[\ln\frac{m^2(\phi,R)}{\mu^2} - \frac{3}{2}\Bigr] \nonumber\\
&\quad + \frac{1}{192\pi^2}\Bigl(\xi - \frac{1}{6}\Bigr)m^2(\phi,R) R \nonumber\\
&\quad \times \Bigl[\ln\frac{m^2(\phi,R)}{\mu^2} - 1\Bigr] + \cdots, \label{eq:DeltaU_curved}
\end{align}
where dots denote higher-order terms in curvature and total derivatives. In the $O(N)$ model analyzed in the main text, the curvature-dependent terms beyond those captured by $m^2(v,\bar{R})$ can be absorbed into the running of $\alpha_R(\mu)$ and $\xi(\mu)$ in the approximation employed, justifying the simplified form (\ref{eq:U1loop}).

\bibliography{taohuang}

\end{document}